\documentclass[12pt]{iopart}
\expandafter\let\csname equation*\endcsname\relax
\expandafter\let\csname endequation*\endcsname\relax
\usepackage{amsmath}
\usepackage{graphicx}

\usepackage{dcolumn}

\usepackage{bm}
\usepackage[normalem]{ulem}
\usepackage{color}
\usepackage{epstopdf}

\newcommand{\bew}{\begin{widetext}}
\newcommand{\ew}{\end{widetext}}

\newcommand{\bx}{\mathbf{x}}

\newcommand{\by}{\mathbf{y}}

\newcommand{\bv}{\mathbf{v}}

\newcommand{\bbr}{\mathbf{r}}
\newcommand{\bQ}{\mathbf{Q}}
\newcommand{\bR}{\mathbf{R}}

\newcommand{\bn}{\mathbf{n}}
\newcommand{\bw}{\mathbf{w}}
\newcommand{\bg}{\mathbf{g}}

\newcommand{\tri}{\triangle}

\newcommand{\bJ}{\mathbf{J}}

\newcommand{\sep}{ \ \ \ , \ \ \ }

\newcommand{\beq}{\begin{equation}}
\newcommand{\eeq}{\end{equation}}
\newcommand{\beqn}{\begin{eqnarray}}
\newcommand{\eeqn}{\end{eqnarray}}
\newcommand{\pp}{\partial}
\newcommand{\dd}{{\rm d}}

\newcommand{\fig}{Fig.\ }

\newcommand{\la}{\langle}

\newcommand{\ra}{\rangle}

\newcommand{\vnab}{{\bf \nabla}}

\begin{document}

\title[Integral Formulae in Motility-Induced Phase Separation]{An Infinite Set of Integral Formulae for Polar, Nematic, and Higher Order Structures at the Interface of Motility-Induced Phase Separation}

\author{Chiu Fan Lee}

\address{Department of Bioengineering, Imperial College London, South Kensington Campus, London SW7 2AZ, U.K.}
\ead{c.lee@imperial.ac.uk}
\vspace{10pt}

\begin{abstract}
Motility-induced phase separation (MIPS) is a purely non-equilibrium phenomenon in which self-propelled particles phase separate without  any attractive interactions. One surprising feature of MIPS is the emergence of polar, nematic, and higher order structures at the  interfacial region, whose underlying physics remains poorly understood. Starting with a model of MIPS in which all many-body interactions are captured by an effective  speed function and an effective pressure function that depend solely on the local  particle density, I derive analytically an infinite set of integral formulae (IF) for the ordering structures at the interface. I then demonstrate that half of these IF are in fact {\it exact} for a wide class of active Brownian particle systems. Finally, I test the IF by applying them to numerical data from direct  particle dynamics simulation and find  that all the IF remain valid to a great extent.
\end{abstract}

%
%
%
%
%
\section{Introduction}
\begin{figure}
		\begin{center}
			\includegraphics[scale=.6]{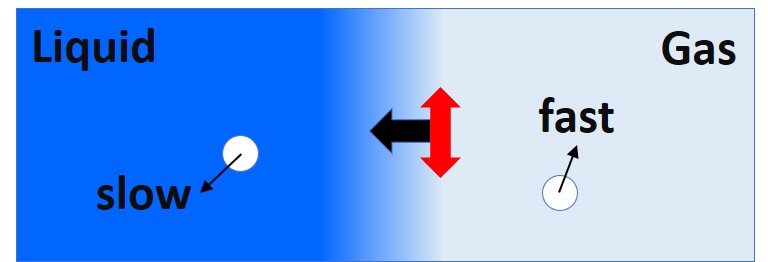}
		\end{center}
		\caption{
			 Schematic of the generic polar-nematic ordering observed at the interface of MIPS. The shade of blue depicts the particle density of the system.	
The particles' orientations are isotropic in the bulks of both liquid and gas phases.
			As one approaches the interface from the liquid (condensed) phase, the average orientation of the particles at the interfacial region becomes predominantly polar (pointing towards the liquid phase as indicated by the black arrow), and then predominantly nematic (i.e., particles' orientations tend be approximately vertical, as indicated by the red double-arrow.)
Higher order tensorial structures beyond the polar and nematic fields also emerge at the interfacial region. Here, using an model in which the effective speed of the particles and the effective local pressure in the system depend solely on the local density, an infinite set of integral formulae is obtained for all these tensorial structures.
		}
		\label{fig:cartoon}
	\end{figure}

The study of active matter is crucial to our understanding of diverse living matter and driven synthetic systems \cite{marchetti_rmp13,needleman_natrev17}.
In addition to being paramount to our quantitative description of various  natural and artificial phenomena, much novel  universal behaviour has also been uncovered in active matter systems in the hydrodynamic limits, which ranges from novel phases \cite{vicsek_prl95,toner_prl95,toner_pre98,toner_prl12,chen_njp18,mahault_prl19,chen_prl20,chen_pre20}
 to  critical phenomena \cite{chen_njp15,mahault_prl18}. 
Besides the  hydrodynamic limits, interesting emergent phenomena also occur in the microscopic and mesoscopic scales. 
In the case of motility-induced phase separation (MIPS) \cite{fily_prl12,redner_prl13,tailleur_prl08,
cates_annrev15,marchetti_curropin16},
one of these emergent phenomena is the 
polar-nematic ordering behaviour   at the liquid-gas interface  (\fig \ref{fig:cartoon})  \cite{lee_softmatter17, patch_softmatt18, solon_njp18,hermann_pre19,omar_pre20}. Understanding this  phenomenon  will be  integral to modelling quantitatively diverse interface-related properties of MIPS that include nucleation  \cite{redner_prl16},  
wetting \cite{wysocki_prl20,neta_softmatt21}, negative interfacial tension \cite{bialke_prl15,paliwal_jcp17,hermann_prl19},  and reverse Ostwald ripening \cite{tjhung_prx18,fausti_a21}.

Although the interfacial  ordering ultimately emerges  from the many-body interactions of the particles, similar  pattern in fact  already  occurs in a system of {\it non-interacting} self-propelled particles (i.e., an ideal active gas) around an impenetrable wall -- particles stuck at the wall tend to point towards it (hence the system is polar), and  particles 
right outside the wall are predominantly moving parallel to the wall (hence nematic), 
 since they constitute particles 
  whose orientations have just rotated   enough to be  pointing away from the wall  \cite{lee_njp13, wagner_jstatmech17}. This revelation
 suggests that the polar-nematic pattern observed in MIPS can be studied using an effective model in which all particle-particle interactions are captured by a  velocity field that depends purely on the  system's local configurations, such as the local particle density. Implementing this task is the goal of this paper. 
 Specifically, I first derive the equation of motion (EOM) of the many-particle  distribution function under  the assumptions that the {\it effective speed} and the {\it effective pressure} depend solely on the local properties of the particle density. I then obtained the steady state equations that describe the polar, nematic, and higher order tensor fields of the system, and from these an infinite set of integral formulae (IF), one for each tensor field. Moreover, I test these IF on published numerical data  obtained from direct particle dynamics simulation \cite{lee_softmatter17}, and demonstrate that the IF remain valid to a remarkable extent. Finally, I showed in the appendix that half of the IF hold {\it exactly} for a wide class of active Brownian particle systems that exhibit MIPS.

\section{Effective model}
In two dimensions, an archetype  of  active systems exhibiting MIPS  consists of a collection of $N$  polar self-propelled  particles in a finite volume $V$, such that the particles interact purely sterically via a short-ranged repulsive potential energy $U$, whose exact form is unimportant. Each particle also exerts a constant active force, $f_a$, that points to a particular orientation denoted by $\hat{\bn} \equiv \cos \Theta \hat{\bx} +\sin \Theta \hat{\by}$, and the angle $\Theta$  itself undergoes diffusion with the rotational diffusion coefficient  $D_{\rm R}$. Furthermore, the particles can potentially perform Brownian motions characterized by the  diffusion coefficient $D_{\rm T}$. Overall, the microscopic equations of motion (EOM) of the system are  therefore
\begin{subequations}
\label{eq:Rtheta}
\begin{align} 
\frac{\dd \bR_i}{\dd t} &= -\frac{1}{\eta} \sum_{j \neq i} \vnab_{\bR_i} U(|\bR_i-\bR_j|) +\frac{f_a}{\eta} \hat{\bn}_i+\sqrt{2D_{\rm T}}\bg^{\rm T}_{i}(t)
\\
\frac{\dd \Theta_i}{\dd t} &= \sqrt{2 D_{\rm R}} g^{\rm R}_i(t)
\ ,
\end{align}
\end{subequations}
where the indices $i,j$ enumerate the particles ($i,j = 1,2, \ldots , N$), $\bR_i(t)$ is the $i$-th particle's position at time $t$, $\eta$ is the damping coefficient in this overdamped system, and 
$\bg^{\rm T}$'s and  $g^{\rm R}$'s are independent Gaussian noise terms with zero means and unit variances, e.g., $\la g^{\rm R}_i(t)\ra=0$ and $\la g^{\rm R}_i(t)g^{\rm R}_j(t')\ra=\delta_{ij}\delta (t-t')$.

Since the particles' identities are irrelevant, we can focus instead on the temporal evolution of the  $N$-particle distribution function:
\beq
\label{eq:psi}
\psi(\bbr, \theta, t) =\left\la\sum_{i=1}^N \delta^2(\bbr-\bR_i(t))\delta(\theta - \Theta_i(t))\right\ra
\ ,
\eeq
where the angular brackets denote averaging over the noises.

As mentioned before, the goal here is to account for all particle-particle interactions and fluctuations effectively through 
a speed function and a pressure function that depend solely  on the local  density: $\rho(\bbr,t)=(2 \pi)^{-1}\int \dd \theta \psi (\bbr, \theta,t)$.  
Specifically, 
the model  EOM of $\psi$ is assumed to be of the form:
\beq
\label{eq:psi_eom}
\pp_t \psi = - \vnab_\bbr \cdot \left( \bv \psi \right) +D_{\rm T} \vnab^2_\bbr \psi +D_{\rm R} \pp_\theta^2 \psi 
\ ,
\eeq
where $\bv$ is the `velocity field' taken to be
\beq
\label{eq:v}
\bv(\bbr, \theta, t) = u(\rho(\bbr ,t))\hat{\bn}(\theta)-\rho(\bbr, t)^{-1} \vnab P(\rho(\bbr, t))
\ ,
\eeq
and  $u$ and $P$ are the effective density-dependent speed and pressure functions, respectively.
Note that the `effective pressure' $P$  does not necessarily correspond to the mechanical pressure in a thermal system, or the Irving-Kirkwood pressure defined for passive systems \cite{Irving_jcp50}. Instead, $P$ should be viewed as a density-dependent scalar function whose spatial gradient contributes to the velocity field as described in Eq.~(\ref{eq:v}). I further note that allowing the effective functions  $u$ and $P$ to depend also on the spatial derivatives of $\rho$ (e.g., $\nabla^2 \rho$) will not alter the key results here. In the Appendix, I will relate the approximation adopted here to a formally exact set of hierarchical EOM.

Given the model equations (\ref{eq:psi_eom},\ref{eq:v}), we are now ready to systematically construct a reduced set of EOM  in which the fields of interest depend on $\bbr$ and $t$, but not $\theta$. Specifically, these fields will be $m$-th rank tensors  of the form: 
\beq
\label{eq:T}
T^{(m)}_{\alpha_1 \cdots \alpha_m}=\frac{1}{2 \pi}\int \dd \theta \hat{n}_{\alpha_1} \hat{n}_{\alpha_2} \cdots \hat{n}_{\alpha_m} (\psi -\rho)
\ , 
\eeq
where the Greek letters enumerate the spatial coordinates and the subtraction of $\rho$ in the integrand ensures that these tensor fields vanish if $\psi$ is  isotropic in $\theta$.
For instance, the {\it unnormalized} polar field ${\bf M}(\bbr,t)$ and nematic field $\bQ$ are the first-rank  and second-rank tensors, respectively: 
 \beq
 M_\alpha = \frac{1}{2\pi}\int \dd \theta \hat{n}_\alpha \psi \sep  Q_{\alpha \beta} = \frac{1}{2\pi}  \left(\int \dd \theta \hat{n}_\alpha \hat{n}_\beta \psi\right) - \frac{ \rho}{2}\delta_{\alpha \beta}
 \ .
 \eeq
These fields are unnormalized since they are not normalized by the density, as opposed to, e.g., the normalized polar field ${\bf M}^{\rm N} =(2\pi \rho)^{-1} \int \dd \theta \hat{\bn} \psi $.

From (\ref{eq:psi_eom},\ref{eq:v}), the EOM of the density and polar fields can be obtained:
\begin{subequations}
\label{eq:main}
\begin{align} 
\pp_t \rho =&  -\pp_\alpha \left[ u M_\alpha -  \pp_\alpha \left(P +D_{\rm T} \rho\right)\right] 
\\
\pp_t M_\alpha =&
-\pp_\beta \left[ u \left( Q_{\alpha \beta} + \frac{ \rho}{2}\delta_{\alpha \beta}\right) -  \frac{M_\alpha}{\rho}\pp_\beta P -D_{\rm T}\pp_\beta M_\alpha\right]
-D_{\rm R} M_\alpha 
\ ,
\end{align} 
\end{subequations}
 where $\pp_\alpha \equiv \pp / \pp r_\alpha$ and repeated indices are summed over. The EOM of higher order tensor fields can be derived similarly, as illustrated next.

	\begin{figure}
		\begin{center}
			\includegraphics[scale=.8]{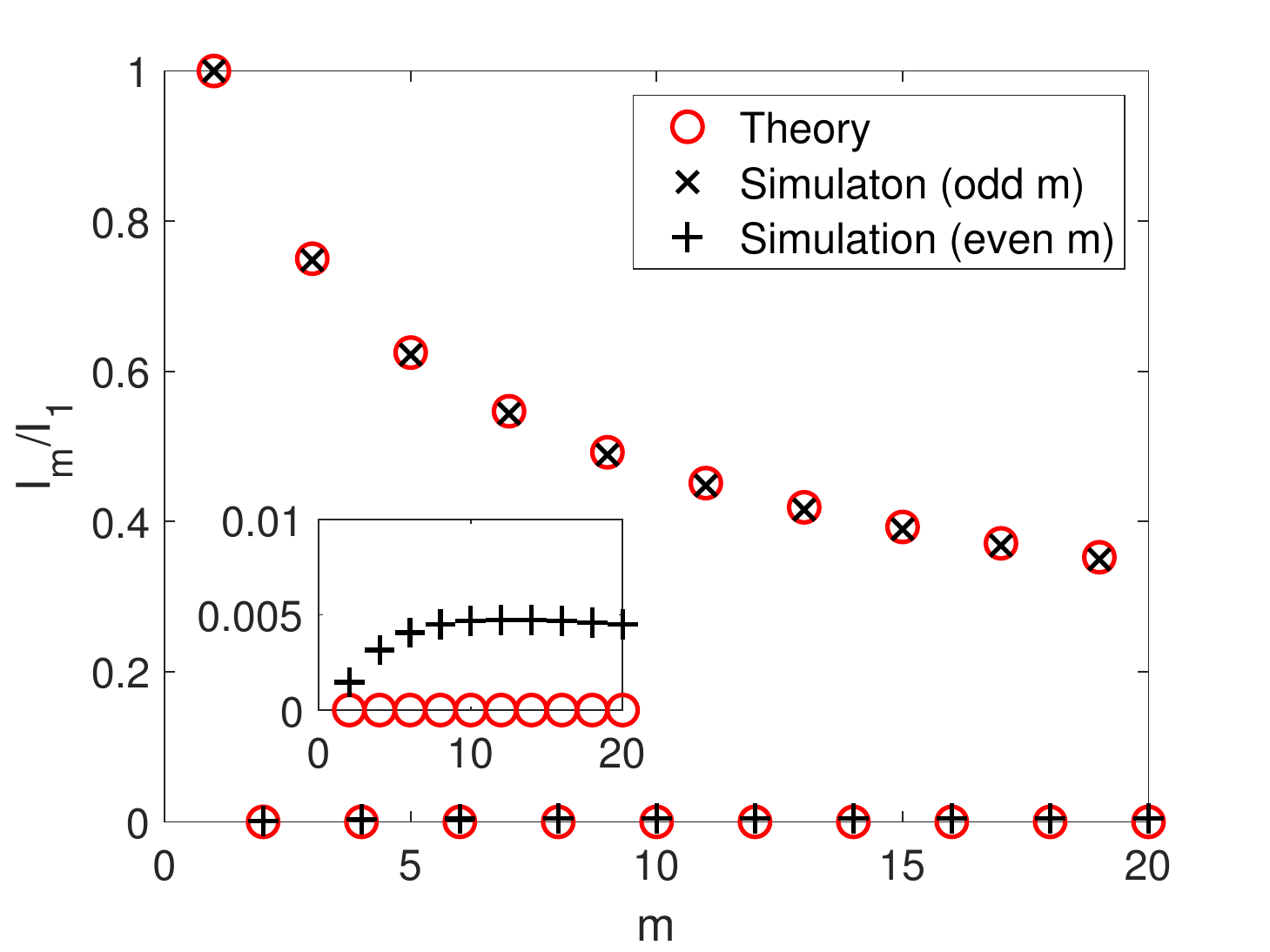}
		\end{center}
		\caption{
		The integrals of the $m$-th rank tensor fields normalized by the integral  of $M_x$ (denoted by $I_1$) (\ref{eq:I}).
The analytical results are denoted by red circles (\ref{eq:int}c,\ref{eq:int2}), and the corresponding  integral values obtained using data from direct particle simulation  are shown by black crosses (odd $m$) and black pluses (even $m$).  The inset plot zooms into the $m$-axis to show the  small deviations of the numerical results from the analytical predictions for even $m$.
		}
		\label{fig1}
	\end{figure}

\section{Integral formulae at the steady state}
I will now focus exclusively on the steady state in the phase separated regime, where there is a single  liquid-gas interface located around $x=0$. By construction, spatial variations occur only along the $x$ direction, and we thus need only to consider tensor fields with all subscripts being $x$, i.e., tensor fields of the form $T^{(m)}_{x \cdots x}$. For later convenience, I will further denote $T^{(m)}_{x \cdots x}$ by $O_m$ when $m$ is odd and by $E_m$ when $m$ is even.  Specifically, from the definitions of the tensor fields (\ref{eq:T}), we have
\beq
\label{eq:def_OE}
O_m = \frac{1}{2\pi} \int \dd \theta \cos^m \theta \psi
\ , \
E_m = \frac{1}{2\pi} \int \dd \theta \cos^m \theta \psi - K_m\rho
\ ,
\eeq
where
\beq
\label{eq:K}
K_m \equiv \frac{(m-1)!!}{2^{m/2-1}m(m/2-1)!}
\ .
\eeq

At the steady state, the equations satisfied by the tensor fields are, from (\ref{eq:psi_eom},\ref{eq:v}):
\begin{subequations}
\label{eq:main_2}
\begin{align} 
D_{\rm R} M_x=&
-\frac{\dd}{\dd x} \left[ u \left( Q_{xx} + \frac{\rho}{2}\right) -\frac{M_x}{\rho} \frac{\dd P}{\dd x} -D_{\rm T} \frac{\dd M_x}{\dd x}\right]
\ ,
\\
4D_{\rm R} Q_{xx} =& 
-\frac{\dd}{\dd x} \left[ u O_{3}-\frac{Q_{xx}+\rho/2}{\rho} \frac{\dd P}{\dd x} -D_{\rm T} \frac{\dd Q_{xx}}{\dd x} \right]
\ ,
\\
D_{\rm R}   \left[m^2 O_m- (m-1)m O_{m-2}\right] =&
-\frac{\dd}{\dd x} \left[ u \left( E_{m+1} + K_{m+1} \rho\right) -\frac{O_m}{\rho} \frac{\dd P}{\dd x} -D_{\rm T} \frac{\dd O_m}{\dd x} \right]
\ , \ \ {\rm for \ odd} \ m> 1 \ ,
\\
D_{\rm R}  \left[m^2 E_{m}- (m-1)m E_{m-2}\right] =& 
-\frac{\dd}{\dd x} \left[ uO_{m+1}
-\frac{E_{m}+K_m \rho}{\rho} \frac{\dd P}{\dd x} -D_{\rm T} \frac{\dd E_m}{\dd x} \right]
\ , \ \ {\rm for \ even} \ m>2 \ .
\end{align} 
\end{subequations}
The L.H.S.~of (\ref{eq:main_2}) follow from the identities below:
\beq
\int \dd \theta \cos^m\theta \pp_\theta^2\psi = \int \dd \theta \left[(m-1)m  \cos^{m-2}\theta  -m^2 \cos^m \theta\right] \psi
\ ,
\eeq
and for the L.H.S. of (\ref{eq:main_2}d), I have additionally used the identity:
\beq
m^2K_m-(m-1)mK_{m-2}=0
\ ,
\eeq
which can be readily derived from the definition of $K_m$ (\ref{eq:K}).

Since $\psi$ is isotropic in both $x$ and $\theta$ deep in the liquid and gas phase, $M_x, Q_{xx}, O_m$, $E_m$, and $\dd P/\dd x$ all vanish when $|x| \gg 0$. Therefore, by integrating both sides of (\ref{eq:main_2}) over the whole $x$ domain, we arrive at the followings:
\begin{subequations}
\label{eq:int}
\begin{align} 
 \int_{-\infty}^\infty \dd x  M_x = K_{2} A =& \frac{A}{2}\ ,
\\
 \int_{-\infty}^\infty \dd x  \left[m^2 O_m- (m-1)m O_{m-2}\right] =& K_{m+1}A \ ,
 \\
 \int_{-\infty}^\infty \dd x   Q_{xx} = \int_{-\infty}^\infty \dd x    E_{m} =&\ 0\ ,
\end{align} 
\end{subequations}
where
\beq
A \equiv \frac{1}{D_{\rm R}}\left[\lim_{x\rightarrow -\infty} u(\rho(x))  \rho(x) -\lim_{x\rightarrow \infty} u(\rho(x))  \rho(x)\right] \ .
\eeq

In fact, one more simplification can be made: starting with (\ref{eq:int}a) and using (\ref{eq:int}b), one can readily prove by mathematical induction that
 \beq
 \label{eq:int2}
  \int_{-\infty}^\infty \dd x  \la O_m \ra=K_{m+1}A \ .
  \eeq
  {\it The integral formulae (IF) expressed in  (\ref{eq:int}c) and (\ref{eq:int2}) are the key results of this paper.}
In the appendix, I will further show that for even $m$, the IF (\ref{eq:int}c) are in fact {\it exact} in a wide class of active Brownian particle systems.

   Introducing the following notation:
 \beq
 \label{eq:I}
I_m = 
\left\{
\begin{array}{ll}
\int \dd x \la O_m \ra\ , & m \ {\rm odd}
\\
\int \dd x \la  E_m \ra\ , & m \ {\rm even}\ ,
\end{array}
\right.
\eeq
the plot of  $I_m/I_1$ vs.~$m$ is shown  in Fig.~\ref{fig1} (red circles). Note in particular the slow decay  of $I_m/I_1$ for odd $m$, which goes asymptotically to 0 as $m^{-1/2}$ (\ref{eq:K}).

%

	\begin{figure}
		\begin{center}
			\includegraphics[scale=.8]{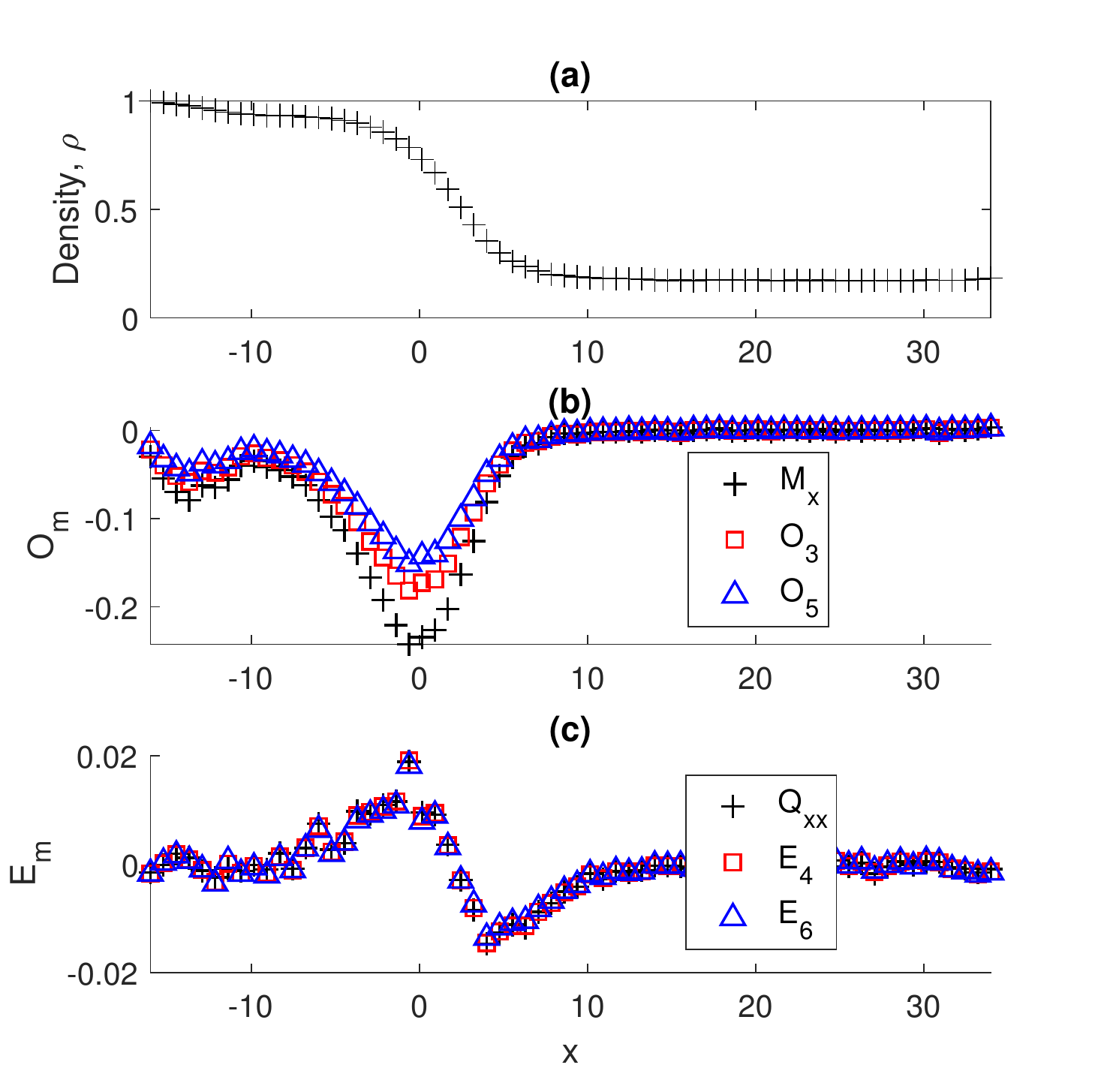}
		\end{center}
		\caption{The variations of the density $\rho$ (a), the odd-rank tensor fields $O_m$ (b), and the even-rank tensor fields $E_m$ (c), with respect to $x$ at the steady state of a system of self-propelled particles with repulsive interactions that exhibits MIPS. The simulation  data used for these curves are from a previous study \cite{lee_softmatter17}.
		}
		\label{fig2}
	\end{figure}

\section{Testing the integral formulae on a microscopic model of MIPS}
As a test of the validity of the IF to realistic active systems exhibiting MIPS, I now re-analyse the numerical results obtained from simulating a microscopic model of MIPS  \cite{lee_softmatter17}. Specifically, the  simulation is on  a system of 300 particles  confined in a 2D channel of height $10\sin (\pi/3)$ and length $50$. 
A periodic boundary condition is used for the vertical direction, a hard wall is placed on the left of the channel, and particles exiting the right wall will stay at the right wall but with its orientation randomized. 
These particles interact via short-ranged repulsive interactions in the form of the 
Weeks–Chandler–Andersen potential \cite{weeks_jcp71}:
\beq
U(r)=
\left\{
\begin{array}{ll}
 \frac{26}{6} \left[ \frac{1}{r^{12}} -\frac{2}{r^{6}}+1 \right]\ , & {\rm if} \ r<1
 \\
 0 \ , & {\rm otherwise} \ .
 \end{array}
 \right.
 \eeq
Other  parameters are: $\eta=1, f_a=100, D_{\rm T}=0$, and $D_{\rm R}=3$. 

At the steady-state, the density and various tensor field profiles are shown in Fig.~\ref{fig2}. Given that the liquid phase is on the left (Fig.~\ref{fig2}a), the odd-rank tensor fields $O_m$ are all negative around the interface (Fig.~\ref{fig2}b), as the particles' orientations are mostly pointing towards the liquid phase (black arrow in Fig.~\ref{fig:cartoon}). 
For the even-rank tensor fields $E_m$ (Fig.~\ref{fig2}c), they first become positive when approaching the interface from deep in the liquid phase, indicating that their orientations are predominantly horizontal, consistent with the orientation of the polar field. 
These fields then subsequently become negative as one exits the interface, indicating that the particles' orientations become predominantly vertical (red double arrow in Fig.~\ref{fig:cartoon}). The ``oscillatory" feature of $E_m$ around the interface is of course necessary in order for the overall integrals of $E_m$ to be zero (\ref{eq:int}c). Note also that while all tensor fields go to zero as $m$ goes to infinity, they apparently do so rather slowly (Fig.~\ref{fig2}(b)-(c)). This  is again consistent with the analytical result that $I_m \sim m^{-1/2}$ for odd $m$. Overall, this slow decays indicates that a quantitative theory focusing on the interfacial region will need to incorporate a large number of higher order tensor fields.
Finally, performing the integrals to this set of data as prescribed in (\ref{eq:I}), the numerical result  (black crosses and pluses in  Fig.~\ref{fig1}) demonstrates that the IF are indeed valid to a great extent.

\section{Summary \& Outlook} 
Starting with a generic microscopic model of active particle system with only repulsive interactions, and adopting the simplifying assumptions that all many-body physics and fluctuations can be  captured by an effective speed function and an effective pressure that depend solely on the local  particle density, I have derived a set of EOM that describes the polar, nematic, and higher order tensor fields of the system. Focusing then on the steady state with a single liquid-gas interface, I obtained an infinite set of IF, one for each tensor field. I then tested these IF on data obtained from direct particle dynamics simulation of a microscopic model of MIPS, and found that the IF remain valid to a high accuracy. Furthermore, I showed in the appendix that half of these IF (even $m$) are in fact {\it exact} for a wide class of active Brownian particle systems. The overall validity of all the IF suggests that the model assumptions are appropriate when studying interfacial properties of MIPS. 

This work opens up a number of interesting future directions: (1) For finite systems, the IF can be made more precise by incorporating the exact boundary conditions at the two limits of the spatial integrals. (2) The IF can  be readily modified to {\it sum rules} for lattice models of MIPS \cite{thompson_jstatmech11, whitelam_jchemphys18,partridge_prl19,shi_prl20}. (3)
 It would be interesting to consider how these IF can be generalized to a system of active particles with alignment interactions. Indeed, the various types of travelling bands observed in polar active matter \cite{csahok_prl95,chate_pre08,gregoire_prl04,bertin_pre06,bertin_jpa09,peruani_prl11,thuroff_prx14,
 schnyder_scirep17,geyer_prx19,nesbitt_njp21,bertrand_a20} are reminiscent of the soliton solutions in the  Korteweg–De Vries equation, which also admits an infinite number of integrals of motion. (4) Finally, the analytical treatment here provides a basis for developing a  quantitative theory that elucidates how the interfacial ordering  impacts upon  other interface-related  phenomena in MIPS, such as the Gibbs-Thomson relation \cite{lee_softmatter17}, wetting \cite{wysocki_prl20,neta_softmatt21}, and reverse Ostwald ripening \cite{tjhung_prx18,fausti_a21}. 
	

\ack
I thank an anonymous referee for their suggestion to look into the {\it exactness} of the IF, which led to the realisation that half of the IF can be shown to be exact from first principles without resorting to the approximation adopted in the main text.

\appendix
\section{Relating the approximation to an exact hierarchical EOM}
In this appendix, I will relate the key approximation made in the main text to a formally exact set of EOM of the tensor fields. To do so, I will
start with the {\it fluctuating}  $N$-particle distribution function:
\beq
\psi^{(f)}(\bbr, \theta, t) =\sum_{i=1}^N \delta^2(\bbr-\bR_i(t))\delta(\theta - \Theta_i(t))
\ ,
\eeq
where the superscript $(f)$ specifies that $\psi^{(f)}$ is a fluctuating quantity (hence different from $\psi$ (\ref{eq:psi})) due to the lack of noise averaging (hence without the angular brackets).

Following standard procedures \cite{tailleur_prl08,cates_epl13,farrell_prl12,dean_jpa96,solon_prl15,solon_njp18},  
the model  EOM of $\psi^{(f)}$ is of the form:
\beq
\label{Aeq:psi_eom}
\pp_t \psi^{(f)} = - \vnab_\bbr \cdot\bJ^{(f)} +D_{\rm T} \vnab^2_\bbr \psi^{(f)} +D_{\rm R} \pp_\theta^2 \psi^{(f)} 
-\vnab_\bbr \cdot \left( \sqrt{2D_{\rm T} \psi^{(f)}} \bg^T\right) - \pp_\theta \left(\sqrt{2D_{\rm R} \psi^{(f)}} g^R\right)
\ ,
\eeq
where $\bJ^{(f)}$ is given by
\beq
\label{Aeq:v}
\bJ^{(f)}(\bbr, \theta,t) = \left[u_0\hat{\bn}(\theta) +\bw^{(f)} (\bbr, t)\right] \psi^{(f)}(\bbr, \theta, t)
\ ,
\eeq
 $u_0\equiv f_a/\eta$ is the constant `active speed', and 
\beq
\bw^{(f)}(\bbr,t) = -\frac{1}{\eta} \int \dd^2 \bbr'\rho^{(f)}(\bbr', t)
\vnab_{\bbr'} U(|\bbr'-\bbr|)
\ ,
\eeq
with $\rho^{(f)}$ being  the `fluctuating' density:
\beq
\rho^{(f)}(\bbr',t)=\frac{1}{2\pi} \int \dd \theta \psi^{(f)}(\bbr', \theta, t)
\ .
\eeq

From (\ref{Aeq:psi_eom},\ref{Aeq:v}), the steady state of the average tensor fields can be obtained similarly as before:
\begin{subequations}
\label{Aeq:main_2}
\begin{align} 
D_{\rm R} M_x =&
-\frac{\dd}{\dd x} \left[ u_0 \left(  Q_{xx} + \frac{ \rho }{2}\right) +
\la M_x^{(f)} w_x^{(f)} \ra -D_{\rm T} \frac{\dd  M_x}{\dd x}\right]
\ ,
\\
4D_{\rm R}  Q_{xx} =& 
-\frac{\dd}{\dd x} \left[ u_0   O_{3}  +\left\la\left(  Q_{xx}^{(f)} + \frac{ \rho^{(f)} }{2}\right) w_x^{(f)} \right\ra-D_{\rm T} \frac{\dd  Q_{xx}}{\dd x} \right]
\ ,
\end{align} 
\end{subequations}
and for odd $m>1$:
\beq
\label{Aeq:oddm}
D_{\rm R}   \left[m^2  O_m- (m-1)m O_{m-2}\right] =
-\frac{\dd}{\dd x} \left[ u_0 \left( E_{m+1} + K_{m+1} \rho\right) +\la O_m^{(f)} w_x^{(f)}\ra  -D_{\rm T} \frac{\dd  O_m }{\dd x} \right]
\ ,
\eeq
and for even $m>2$:
\beq
\label{Aeq:evenm}
D_{\rm R}  \left[m^2  E_{m} - (m-1)m  E_{m-2}\right] =
-\frac{\dd}{\dd x} \left[ u_0  O_{m+1}
+\la ( E_{m}^{(f)} +K_m \rho^{(f)}) w_x^{(f)} \ra-D_{\rm T} \frac{\dd  E_m}{\dd x} \right]
\ .
\eeq

The above steady state equations are similar to those in the main text (\ref{eq:main_2}), except for two crucial differences: (1) the active speed $u_0$ is a constant here instead of being dependent on the density $\rho$, and (2) the previous effective pressure gradient terms are now replaced by the correlation functions between $w_x^{(f)}$ and the associated tensor fields (e.g., of the form $\la O_m^{(f)} w_x^{(f)}\ra$, etc).

The success of the approximation scheme employed in the main text thus indicates that:
\beq
\label{Aeq:u}
\lim_{x\rightarrow \pm \infty} u(\rho) \simeq u_0 +  \lim_{x\rightarrow \pm \infty}\frac{\la O_m^{(f)} w_x^{(f)} \ra}{K_{m+1} \rho} \sep {\rm for \ all \ odd} \ m
\ ,
\eeq
and 
\beq
\label{Aeq:E}
\lim_{x\rightarrow  \infty}\la ( E_{m}^{(f)} +K_m \rho^{(f)}) w_x^{(f)} \ra \simeq \lim_{x\rightarrow -\infty}\la ( E_{m}^{(f)} +K_m \rho^{(f)}) w_x^{(f)} \ra  \sep {\rm for \ all \ even} \ m
\ .
\eeq

Let us first look at  $\la E_m^{(f)} w_x^{(f)} \ra$ for even $m$,  which is given by 
\beq
\label{Aeq:I}
-\frac{1}{(2\pi)^2 \eta} 
\int \dd^2 \bbr'\dd \theta\cos^m \theta  
\left[\pp_{x'} U(|\bbr'-\bbr|)\right]
\la\psi^{(f)} (\bbr, \theta)
 \rho^{(f)}(\bbr')\rangle\ .
\eeq
Deep in the bulks of the liquid and gas phases, we expect that that the correlation  $\la\psi^{(f)} (\bbr, \theta)
 \rho^{(f)}(\bbr')\rangle $ cannot distinguish left from right. Hence 
\beq
\label{Aeq:sym}
\la\psi^{(f)} (x,y, \theta)
 \rho^{(f)}(x+\tri x ,y')\rangle = \la\psi^{(f)} (x,y, \theta+\pi)
 \rho^{(f)}(x-\tri x,y')\rangle 
 \ .
 \eeq
Because of this symmetry, the integral in (\ref{Aeq:I}) is exactly zero, i.e.,
\beq
\la E_m^{(f)} w_x^{(f)} \ra =0 
\ ,
\eeq
for all even $m$. As a result, Eqs.~(\ref{Aeq:E}) are always satisfied and the IF are exact for even $m$.

Let us now focus on odd $m$. In other for the relations in Eq.~(\ref{Aeq:u}) to be satisfied deep in the liquid and gas phases, it is clear that $\la O_m^{(f)} w_x^{(f)} \ra$ has to be proportional to $K_{m+1}$ (so that the relations (\ref{Aeq:u}) are valid for all odd $m$), which implies that 
 \beq
 \label{Aeq:H}
 -\frac{1}{2\pi \eta} 
\lim_{x\rightarrow \pm \infty}\int \dd^2 \bbr'  
\left[\pp_{x'} U(|\bbr'-\bbr|)\right]
\la\psi^{(f)} (\bbr, \theta)
 \rho^{(f)}(\bbr')\rangle   \simeq -H_\pm \cos \theta 
 \ ,
\eeq
where $H_{\pm}$ are constants at the $x \rightarrow \pm \infty$ limits. While the form in (\ref{Aeq:H}) expectedly satisfies the symmetry in (\ref{Aeq:sym}), I cannot show that the integrals on the L.H.S.\ are exactly proportional to $\cos \theta$, and their demonstrations   from first principles will be a very interesting future direction.

\section*{References}


\providecommand{\newblock}{}

\end{document}